\documentclass{article}
\usepackage{amssymb}

\input{tcilatex}
\begin{document}

\author{Emilio Santos \and Universidad de Cantabria. Santander. Spain. \and Email 
\TEXTsymbol{<}emilio.santos@unican.es\TEXTsymbol{>}.}
\title{Local realistic interpretation of entangled photon pairs in the Weyl-Wigner
formalism}
\date{February, 3, 2020 }
\maketitle

\begin{abstract}
A polarization correlation experiment with two maximally entangled photons
created by spontaneous parametric down-conversion is studied in the
Weyl-Wigner formalism, that reproduces the quantum predictions. An
interpretation is proposed in terms of stochastic processes assuming that
the quantum vacuum fields are real. This proves that local realism is
compatible with the violation of Bell inequalities, thus rebutting the claim
that local realism has been refuted by entangled photon experiments.
Entanglement appears as a correlation between fluctuations of a signal field
and vacuum fields.
\end{abstract}

\section{ The empirical refutation of Bell\'{}s local realism}

In 2015 experiments were reported showing for the first time the
loophole-free violation of a Bell inequality\cite{Shalm},\cite{Giustina}.
The result has been interpreted as the ``death by experiment for local
realism'', this being the hypothesis that ``the world is made up of real
stuff, existing in space and changing only through local interactions ...
about the most intuitive scientific postulate imaginable''\cite{Wiseman}.
This statement, and many similar ones, emphasize both the relevance of 
\textit{local realism }for our understanding of the physical world and the
fact that it has been \textit{refuted empirically}. Nevertheless it is worth
studying the possibility of a loophole in the empirical refutation via a new
definition of locality weaker than Bell\'{}s. In this article I search for
such a weak locality, compatible with the said experiments\cite{Shalm},\cite
{Giustina}, that involved photon pairs entangled in polarization produced
via spontaneous parametric down conversion. Thus I will analyze such
experiments using the Weyl-Wigner formalism of quantum optics, rather than
the more usual Hilbert-space formalism. Previously I revisit briefly the
origin and meaning of the Bell inequalities\cite{SantosEJP}.

Bell defined ``local hidden variables'' model, later named ``local
realistic'', to be any model of an experiment where the results of all
measurements may be interpreted according to the formulas

\begin{eqnarray}
\left\langle A\right\rangle &=&\int \rho \left( \lambda \right) d\lambda
M_{A}\left( \lambda ,A\right) ,\left\langle B\right\rangle =\int \rho \left(
\lambda \right) d\lambda M_{B}\left( \lambda ,B\right) ,  \nonumber \\
\left\langle AB\right\rangle &=&\int \rho \left( \lambda \right) d\lambda
M_{A}\left( \lambda ,A\right) M_{B}\left( \lambda ,B\right) ,  \label{bell}
\end{eqnarray}
where $\lambda \in \Lambda $ is one or several random\ (``hidden'')
variables, $\left\langle A\right\rangle ,\left\langle B\right\rangle $ and $%
\left\langle AB\right\rangle $ being the expectation values of the results
of measuring the observables $A,B$ or their product $AB$, respectively. Here
we will consider that the observables correspond to detection, or not, of
some signals (e. g. photons) by two parties named Alice and Bob, attaching
the values 1 or 0 to these two possibilities. In this case $\left\langle
A\right\rangle ,\left\langle B\right\rangle $ correspond to the single and $%
\left\langle AB\right\rangle $ to the coincidence detection rates
respectively. The following mathematical conditions are assumed 
\begin{equation}
\rho \left( \lambda \right) \geq 0,\int \rho \left( \lambda \right) d\lambda
=1,M_{A}\left( \lambda ,A\right) \in \left\{ 0,1\right\} ,M_{B}\left(
\lambda ,B\right) \in \left\{ 0,1\right\} .  \label{bell1}
\end{equation}
Eqs.$\left( \ref{bell1}\right) $ corresponds to a ``deterministic model''
where the statistical aspects derive from the probabilistic nature of the
hidden random variables $\left\{ \lambda \right\} .$ More general models may
be constructed where the whole interval $\left[ 0,1\right] $ is substituted
for $\left\{ 0,1\right\} $ in eq.$\left( \ref{bell1}\right) .$ A constraint
of locality is included, namely $M_{A}\left( \lambda ,A\right) $ should be
independent of the choice of the observable $B$, $M_{B}\left( \lambda
,B\right) $ independent of $A$ and $\rho \left( \lambda \right) $
independent of both $A$ and $B$\cite{Bell}. From these conditions it is
possible to derive empirically testable (Bell) inequalities\cite{CH}, \cite
{Eberhard}. The tests are most relevant if the measurements performed by
Alice and Bob are spacially separated in the sense of relativity theory.

For experiments measuring polarization correlation of photon pairs the
Clauser-Horne inequality\cite{CH} may be written

\begin{equation}
\left\langle \theta _{1}\right\rangle +\left\langle \phi _{1}\right\rangle
\geq \left\langle \theta _{1}\phi _{1}\right\rangle +\left\langle \theta
_{1}\phi _{2}\right\rangle +\left\langle \theta _{2}\phi _{1}\right\rangle
-\left\langle \theta _{2}\phi _{2}\right\rangle ,  \label{CH}
\end{equation}
where $\theta _{j}$ stands for the observable ``detection of a photon with
the Alice detector in front of a polarizer at angle $\theta _{j}$''.
Similarly $\phi _{k}$ for Bob detector. For simplicity in this paper I will
study the case of maximally entangled photons, although in the mentioned
experiments\cite{Shalm},\cite{Giustina} the photon pairs had partial
entanglement, which made the experiments easier. I will present a local
model that predicts the following single and coincidence rates by Alice and
Bob 
\begin{eqnarray}
\left\langle \theta _{j}\right\rangle &=&\left\langle \phi _{k}\right\rangle
=\frac{1}{2}K,  \nonumber \\
\left\langle \theta _{j}\phi _{k}\right\rangle &=&\frac{1}{2}K\cos
^{2}\left( \theta _{j}-\phi _{k}\right) =\frac{1}{4}K\left[ 1+\cos \left(
2\theta _{j}-2\phi _{k}\right) \right] ,  \label{pred}
\end{eqnarray}
where $K$ is a constant that depends on the particular experimental setup.
It is easy to check that the prediction eq.$\left( \ref{pred}\right) $
violates the inequality eq.$\left( \ref{CH}\right) $ for some choices of
angles. For instance the choice 
\[
\theta _{1}=\frac{\pi }{4},\phi _{1}=\frac{\pi }{8},\theta _{2}=0,\phi _{2}=%
\frac{3\pi }{8}, 
\]
violates the inequality eq.$\left( \ref{CH}\right) $ leading to 
\[
K\ngtr \frac{1}{2}\left( 1+\sqrt{2}\right) K\simeq 1.207K. 
\]
So far we have assumed ideal detectors, for real detectors the predicted
single rates $\left\langle \theta _{j}\right\rangle $ and $\left\langle \phi
_{k}\right\rangle $ should be multiplied times the detection efficiencies $%
\eta _{A}$ and $\eta _{B},$ respectively, and the coincidence rate $%
\left\langle \theta _{j}\phi _{k}\right\rangle $ times $\eta _{A}\eta _{B},$
whence the empirical violation of the inequality eq.$\left( \ref{CH}\right) $
would require high detection efficiencies, that is 
\[
\eta _{A}+\eta _{B}<(1+\sqrt{2})\eta _{A}\eta _{B}\Rightarrow \eta >0.828%
\text{ if }\eta _{A}=\eta _{B}=\eta . 
\]
Experiments with some non-maximal entanglement need only $\eta >2/3$ \cite
{Eberhard}, that was the reason for using such entanglement in the actual
experiments \cite{Shalm}, \cite{Giustina}.

In the following sections I shall shortly review the treatment within the
Weyl-Wigner formalism of the polarization correlation measurement of two
maximally entangled photons produced via spontaneous parametric down
conversion (SPDC). Thus I continue a theoretical interpretation of SPDC
experiments within the WW formalism in the Heisenberg picture, that was
initiated in the nineties of the past century \cite{C1} - \cite{C11}. In
many of those early studies the approach was heuristic and one of the
purposes of this article is to provide a more formal foundation. The WW
formalism suggests an intuitive picture for photon entanglement and the
interpretation of SPDC experiments in terms of random variables and
stochastic processes.

\section{The Weyl-Wigner formalism in quantum optics}

\subsection{Definition}

The WW formalism was developped for non-relativistic quantum mechanics,
where the basic observables involved are positions, $\hat{x}_{j},$ and
momenta, $\hat{p}_{j},$ of the particles\cite{Weyl}, \cite{Wigner}, \cite
{Groenewold}, \cite{Moyal}, \cite{Scully}, \cite{Zachos}. It may be
trivially extended to quantum optics provided we interpret $\hat{x}_{j}$ and 
$\hat{p}_{j}$ to be the sum and the difference of the creation, $\hat{a}%
_{j}^{\dagger },$ and annihilation, $\hat{a}_{j},$ operators of the $j$
normal mode of the radiation. That is

\begin{eqnarray}
\hat{x}_{j} &\equiv &\frac{c}{\sqrt{2}\omega _{j}}\left( \hat{a}_{j}+\hat{a}%
_{j}^{\dagger }\right) ,\hat{p}_{j}\equiv \frac{i 
\rlap{\protect\rule[1.1ex]{.325em}{.1ex}}h%
\omega _{j}}{\sqrt{2}c}\left( \hat{a}_{j}-\hat{a}_{j}^{\dagger }\right) 
\nonumber \\
&\Rightarrow &\hat{a}_{j}=\frac{1}{\sqrt{2}}\left( \frac{\omega _{j}}{c}\hat{%
x}_{j}+\frac{ic}{
\rlap{\protect\rule[1.1ex]{.325em}{.1ex}}h%
\omega }\hat{p}_{j}\right) ,\hat{a}_{j}^{\dagger }=\frac{1}{\sqrt{2}}\left( 
\frac{\omega _{j}}{c}\hat{x}_{j}-\frac{ic}{
\rlap{\protect\rule[1.1ex]{.325em}{.1ex}}h%
\omega _{j}}\hat{p}_{j}\right) .  \label{0}
\end{eqnarray}
Here $
\rlap{\protect\rule[1.1ex]{.325em}{.1ex}}h%
$ is Planck constant, $c$ the velocity of light and $\omega _{j}$ the
frequency of the normal mode. In the following I will use units $
\rlap{\protect\rule[1.1ex]{.325em}{.1ex}}h%
=c=1$. For the sake of clarity I shall represent the operators in a Hilbert
space with a `hat', e. g. $\hat{a}_{j},\hat{a}_{j}^{\dagger }$ and the
amplitudes in the WW formalism without `hat', e. g. $a_{j},a_{j}^{*}.$

The connection with the Hilbert-space formalism is made via the Weyl
transform as follows. For any trace class operator $\hat{M}$ of the former
we define its Weyl transform to be a function of the field operators $%
\left\{ \hat{a}_{j},\hat{a}_{j}^{\dagger }\right\} $, that is 
\begin{eqnarray*}
W_{\hat{M}} &=&\frac{1}{(2\pi ^{2})^{n}}\prod_{j=1}^{n}\int_{-\infty
}^{\infty }d\lambda _{j}\int_{-\infty }^{\infty }d\mu _{j}\exp \left[
-2i\lambda _{j}\mathrm{Re}a_{j}-2i\mu _{j}\mathrm{Im}a_{j}\right] \\
&&\times Tr\left\{ \hat{M}\exp \left[ i\lambda _{j}\left( \hat{a}_{j}+\hat{a}%
_{j}^{\dagger }\right) +i\mu _{j}\left( \hat{a}_{j}-\hat{a}_{j}^{\dagger
}\right) \right] \right\} .
\end{eqnarray*}
The transform is \textit{invertible} that is 
\begin{eqnarray*}
\hat{M} &=&\frac{1}{(2\pi ^{2})^{2n}}\prod_{j=1}^{n}\int_{-\infty }^{\infty
}d\lambda _{j}\int_{-\infty }^{\infty }d\mu _{j}\exp \left[ i\lambda
_{j}\left( \hat{a}_{j}+\hat{a}_{j}^{\dagger }\right) +i\mu _{j}\left( \hat{a}%
_{j}-\hat{a}_{j}^{\dagger }\right) \right] \\
&&\times \prod_{j=1}^{n}\int_{-\infty }^{\infty }d\mathrm{Re}%
a_{j}\int_{-\infty }^{\infty }d\mathrm{Im}a_{j}W_{\hat{M}}\left\{
a_{j},a_{j}^{*}\right\} \exp \left[ -2i\lambda _{j}\mathrm{Re}a_{j}-2i\mu
_{j}\mathrm{Im}a_{j}\right] .
\end{eqnarray*}
The transform is\textit{\ linear}, that is if $f$ is the transform of $\hat{f%
}$ and $g$ the transform of $\hat{g}$, then the transform of $\hat{f}$ +$%
\hat{g} $ is $f+g$.

It is standard wisdom that the WW formalism is unable to provide any
intuitive picture of the quantum phenomena. The reason is that the Wigner
function, that represents the quantum states, is not positive definite in
general and therefore cannot be interpreted as a probability distribution
(of positions and momenta in quantum mechanics, or field amplitudes in
quantum optics). However we shall see that in quantum optics the formalism
in the Heisenberg representation, where the evolution goes in the field
amplitudes, allows the interpretation of the experiments using the Wigner
function only for the vacuum state, that is positive definite.

The use of the WW formalism in quantum optics has the following features in
comparison with the Hilbert-space formalism:

1. It is just quantum optics, therefore the predictions for experiments are
the same.

2. The calculations using the WW formalism are generally no more involved
than the corresponding ones in Hilbert space, and sometimes they are easier
because no problem of non-commutativity arises.

3. The formalism suggests a physical picture in terms of random variables
and stochastic processes. In particular the counterparts of creation and
annihilation operators look like random amplitudes.

Here we shall use the formalism in the Heisenberg picture, where the
evolution appears in the observables. On the other hand the concept of
photon does not appear in the WW formalism.

\subsection{Properties}

All properties of the WW transform in particle systems may be translated to
quantum optics via eqs.$\left( \ref{0}\right) .$ The transform allows
getting a function of (c-number) amplitudes for any trace-class operator (
e. g. any function of the creation and annihilation operators of `photons').
In particular we may get the (Wigner) function corresponding to any quantum
state. For instance the vacuum state, represented by the density matrix $%
\left| 0\rangle \langle 0\right| ,$ is associated to the following Wigner
function 
\begin{equation}
W_{0}=\prod_{j}\frac{2}{\pi }\exp \left( -2\left| a_{j}\right| ^{2}\right) .
\label{1}
\end{equation}
This function may be interpreted as a (positive) probability distribution.
Hence the picture that emerges is that the quantum vacuum of the
electromagnetic field (also named \textit{zeropoint field, ZPF}) consists of
stochastic fields with a probability distribution independent for every
mode, having a Gaussian distribution with mean energy $\frac{1}{2} 
\rlap{\protect\rule[1.1ex]{.325em}{.1ex}}h%
\omega $ per mode.

Similarly there are functions associated to the observables. For instance
the following Weyl transforms are obtained 
\begin{eqnarray}
\hat{a}_{j} &\leftrightarrow &a_{j},\hat{a}_{j}^{\dagger }\leftrightarrow
a_{j}^{*},\frac{1}{2}\left( \hat{a}_{j}^{\dagger }\hat{a}_{j}+\hat{a}_{j}%
\hat{a}_{j}^{\dagger }\right) \leftrightarrow a_{j}a_{j}^{*}=\left|
a_{j}\right| ^{2},  \nonumber \\
\hat{a}_{j}^{\dagger }\hat{a}_{j} &=&\frac{1}{2}\left( \hat{a}_{j}^{\dagger }%
\hat{a}_{j}+\hat{a}_{j}\hat{a}_{j}^{\dagger }\right) +\frac{1}{2}\left( \hat{%
a}_{j}^{\dagger }\hat{a}_{j}-\hat{a}_{j}\hat{a}_{j}^{\dagger }\right)
\leftrightarrow \left| a_{j}\right| ^{2}-\frac{1}{2},  \nonumber \\
\left( \hat{a}_{j}^{\dagger }+\hat{a}_{j}\right) ^{n} &\leftrightarrow
&\left( a_{j}+a_{j}^{*}\right) ^{n},\left( \hat{a}_{j}^{\dagger }-\hat{a}%
_{j}\right) ^{n}\leftrightarrow \left( a_{j}-a_{j}^{*}\right) ^{n},n\text{
an integer.}  \label{2}
\end{eqnarray}
I stress that the quantities $a_{j}$ and $a_{j}^{*}$ are c-numbers and
therefore they commute with each other. The former eqs.$\left( \ref{2}%
\right) $ mean that in expressions \textit{linear in creation and/or
annihilation operator} the Weyl transform just implies ``\textit{removing
the hats}''. However this is not the case in nonlinear expressions in
general. In fact from the latter two eqs.$\left( \ref{2}\right) $ plus the
linearity property it follows that for a product in the WW formalism the
Hilbert space counterpart is 
\begin{equation}
a_{j}^{k}a_{j}^{*^{l}}\leftrightarrow (\hat{a}_{j}^{k}\hat{a}_{j}^{\dagger
l})_{sym},  \label{2b}
\end{equation}
where the subindex $sym$ means writing the product with all possible
orderings and dividing for the number of terms. Hence the WW field
amplitudes corresponding to products of field operators may be obtained
putting the operators in symmetrical order via the commutation relations.
Particular instances that will be useful latter are the following 
\begin{eqnarray}
\hat{a}_{j}^{\dagger }\hat{a}_{j} &\rightarrow &\left| a_{j}\right| ^{2}-%
\frac{1}{2},\hat{a}_{j}\hat{a}_{j}^{\dagger }\rightarrow \left| a_{j}\right|
^{2}+\frac{1}{2},\hat{a}_{j}{}^{2}\rightarrow a_{j}^{2},\hat{a}_{j}^{\dagger
2}\rightarrow \hat{a}_{j}^{*2}  \nonumber \\
\hat{a}_{j}^{\dagger }\hat{a}_{j}\hat{a}_{j}^{\dagger }\hat{a}_{j}
&\rightarrow &\left| a_{j}\right| ^{4}-\left| a_{j}\right| ^{2},\hat{a}_{j}%
\hat{a}_{j}^{\dagger }\hat{a}_{j}\hat{a}_{j}^{\dagger }\rightarrow \left|
a_{j}\right| ^{4}+\left| a_{j}\right| ^{2},  \label{b2} \\
\hat{a}_{j}^{\dagger }\hat{a}_{j}^{\dagger }\hat{a}_{j}\hat{a}_{j}
&\rightarrow &\left| a_{j}\right| ^{4}-2\left| a_{j}\right| ^{2}+\frac{1}{2},%
\hat{a}_{j}\hat{a}_{j}\hat{a}_{j}^{\dagger }\hat{a}_{j}^{\dagger
}\rightarrow \left| a_{j}\right| ^{4}+2\left| a_{j}\right| ^{2}+\frac{1}{2}.
\nonumber
\end{eqnarray}

Other properties may be easily obtained from well known results of the
standard Weyl-Wigner formalism in particle quantum mechanics. I will present
them omitting the proofs.

\textit{Expectation values} may be calculated in the WW formalism as
follows. In the Hilbert-space formalism they read $Tr(\hat{\rho}\hat{M})$,
or in particular $\langle \psi \mid \hat{M}\mid \psi \rangle ,$ whence the
translation to the WW formalism is obtained taking into account that the
trace of the product of two operators becomes 
\[
Tr(\hat{\rho}\hat{M})=\int W_{\hat{\rho}}\left\{ \hat{a}_{j},\hat{a}%
_{j}^{\dagger }\right\} W_{\hat{M}}\left\{ \hat{a}_{j},\hat{a}_{j}^{\dagger
}\right\} \prod_{j}d\mathrm{Re}a_{j}d\mathrm{Im}a_{j}. 
\]
That integral is the WW counterpart of the trace operation in the
Hilbert-space formalism. Particular instances are the following expectations
that will be of interest later on 
\begin{eqnarray}
\left\langle \left| a_{j}\right| ^{2}\right\rangle &\equiv &\int d\Gamma
W_{0}\left| a_{j}\right| ^{2}=\frac{1}{2},\left\langle
a_{j}^{n}a_{j}^{*m}\right\rangle =0\text{ if }n\neq m.  \nonumber \\
\left\langle 0\left| \hat{a}_{j}^{\dagger }\hat{a}_{j}\right| 0\right\rangle
&=&\int d\Gamma (a_{j}^{*}a_{j}-\frac{1}{2})W_{0}=0,  \nonumber \\
\left\langle 0\left| \hat{a}_{j}\hat{a}_{j}^{\dagger }\right| 0\right\rangle
&=&\int d\Gamma (\left| a_{j}\right| ^{2}+\frac{1}{2})W_{0}=2\left\langle
\left| a_{j}\right| ^{2}\right\rangle =1,  \label{3a} \\
\left\langle \left| a_{j}\right| ^{4}\right\rangle &=&1/2,\left\langle
\left| a_{j}\right| ^{n}\left| a_{k}\right| ^{m}\right\rangle =\left\langle
\left| a_{j}\right| ^{n}\right\rangle \left\langle \left| a_{k}\right|
^{m}\right\rangle \text{ if }j\neq k.  \nonumber
\end{eqnarray}
where $W_{0}$ is the Wigner function of the vacuum, eq.$\left( \ref{1}%
\right) $. This means that in the WW formalism the field amplitude $a_{j}$
(coming from the vacuum) behaves like a complex random variable with
Gaussian distribution and mean square modulus $\left\langle \left|
a_{j}\right| ^{2}\right\rangle =1/2.$ I point out that the integral for any
mode not entering in the function $M\left( \left\{ a_{j},a_{j}^{*}\right\}
\right) $ gives unity in the integration due to the normalization of the
Wigner function eq.$\left( \ref{1}\right) $. An important consequence of eq.$%
\left( \ref{3a}\right) $ is that \textit{normal (antinormal) ordering of
creation and annihilation operators in the Hilbert space formalism becomes
subtraction (addition) of 1/2 in the WW formalism. The normal ordering rule
is equivalent to subtracting the vacuum contribution.}

\subsection{Evolution}

In the Heisenberg picture of the Hilbert-space formalism the density matrix
is fixed and any observable, say $\hat{M}$, evolves according to 
\[
\frac{d}{dt}\hat{M}=i\left( \hat{H}\hat{M}-\hat{M}\hat{H}\right) ,\hat{M}=%
\hat{M}\left( t\right) , 
\]
where $\hat{H}$ is the Hamiltonian. Translated to the WW formalism this
evolution of the observables is given by the Moyal equation with the sign
changed. The standard Moyal equation applies to the evolution of the Wigner
function, that represents a quantum state being the counterpart of the
density matrix in the Schr\"{o}dinger picture of the Hilbert space
formalism. Thus in the WW formalism we have 
\begin{eqnarray}
\frac{\partial W_{\hat{M}}}{\partial t} &=&2\{\sin \left[ \frac{1}{4}\left( 
\frac{\partial }{\partial \mathrm{Re}a_{j}^{\prime }}\frac{\partial }{%
\partial \mathrm{Im}a_{j}^{\prime \prime }}-\frac{\partial }{\partial 
\mathrm{Im}a_{j}^{\prime }}\frac{\partial }{\partial \mathrm{Re}%
a_{j}^{\prime \prime }}\right) \right]  \nonumber \\
&&\times W_{\hat{M}}\left\{ a_{j}^{\prime },a_{j}^{*\prime },t\right\}
H\left( a_{j}^{\prime \prime },a_{j}^{*\prime \prime }\right) _{a_{j}},
\label{Moyal}
\end{eqnarray}
where $\left\{ {}\right\} _{a_{j}}$ means making $a_{j}^{\prime
}=a_{j}^{\prime \prime }=a_{j}$ and $a_{j}^{*\prime }=a_{j}^{*\prime \prime
}=a_{j}^{*}$ after performing the derivatives.

A simple example is the free evolution of the field amplitude of a single
mode. The Hamiltonian in the WW formalism may be trivially obtained
translating the Hamiltonian of the Hilbert-space formalism, that is 
\begin{eqnarray*}
\hat{H}_{free} &=&\omega _{j}\hat{a}_{j}^{\dagger }\hat{a}_{j}=\frac{1}{2}%
\omega _{j}(\hat{a}_{j}^{\dagger }\hat{a}_{j}+\hat{a}_{j}\hat{a}%
_{j}^{\dagger })-\frac{1}{2}\omega _{j} \\
&\rightarrow &H_{free}=\omega _{j}(\left| a_{j}\right| ^{2}-\frac{1}{2}%
)=\omega _{j}\left[ (\mathrm{Re}a_{j})^{2}+(\mathrm{Im}a_{j})^{2}-\frac{1}{2}%
\right] ,
\end{eqnarray*}
where we have taken the first eq.$\left( \ref{3a}\right) $ into account.
This leads to 
\begin{equation}
\frac{d}{dt}a_{j}=\frac{1}{2}\omega _{j}\left[ 2(\mathrm{Im}a_{j})-2\left( 
\mathrm{Re}a_{j}\right) i\right] =-i\omega _{j}a_{j}\Rightarrow a_{j}\left(
t\right) =a_{j}\left( 0\right) \exp \left( -i\omega _{j}t\right)  \label{a}
\end{equation}

Another example is the down-conversion process in a nonlinear crystal.
Avoiding a detailed study of the physics inside the crystal\cite{Dechoum}, 
\cite{MS} we shall study a single mode problem with the model Hamiltonian%
\cite{Ou} 
\begin{equation}
\hat{H}_{I}=A\hat{a}_{s}^{\dagger }\hat{a}_{i}^{\dagger }\exp \left(
-i\omega _{P}t\right) +A^{*}\hat{a}_{s}\hat{a}_{i}\exp \left( i\omega
_{P}t\right) ,  \label{40}
\end{equation}
when the laser is treated as classically prescribed, undepleted and
spatially uniform field of frequency $\omega _{P}.$ The parameter $A$ is
proportional to the pump amplitude and the nonlinear susceptibility. In the
WW formalism this Hamiltonian becomes (see eqs.$\left( \ref{2}\right) )$%
\[
H_{I}=Aa_{s}^{*}a_{i}^{*}\exp \left( -i\omega _{P}t\right)
+A^{*}a_{s}a_{i}\exp \left( i\omega _{P}t\right) , 
\]
whence taking eqs.$\left( \ref{Moyal}\right) $ and $\left( \ref{a}\right) $
into account we have 
\begin{eqnarray}
\frac{d}{dt}a_{s} &=&-i\omega _{s}a_{s}-iAa_{i}^{*}\exp \left( -i\omega
_{P}t\right) ,  \label{41} \\
\frac{d}{dt}a_{i} &=&-i\omega _{i}a_{i}-iAa_{s}^{*}\exp \left( -i\omega
_{P}t\right) .  \nonumber
\end{eqnarray}
We shall assume that the vacuum field $a_{s}$ evolves as in eq.$\left( \ref
{a}\right) $ before entering the crystal and then according to eqs.$\left( 
\ref{41}\right) $ inside the crystal, that is during the time $T$ needed to
cross it. In order to get the radiation intensity to second order in $%
AT\equiv C$ (see below section 2.4) we must solve these two coupled
equations also to second order. After some algebra this leads to 
\begin{eqnarray}
a_{s}(t) &=&\left( 1+\frac{1}{2}\left| C\right| ^{2}\right) a_{s}(0)\exp
\left( -i\omega _{s}t\right) -iCa_{i}^{*}\left( 0\right) \exp \left[ i\left(
\omega _{i}-\omega _{P}\right) t\right]  \nonumber \\
&=&[\left( 1+\frac{1}{2}\left| C\right| ^{2}\right)
a_{s}(0)-iCa_{i}^{*}\left( 0\right) ]\exp \left( -i\omega _{s}t\right) ,
\label{42}
\end{eqnarray}
and the latter equality takes the `energy conservation' into account (that
in the WW formalism looks like a condition of frequency matching, $\omega
_{P}=\omega _{s}+\omega _{i}$, with no reference to photon energies).

Eq.$\left( \ref{42}\right) $ gives the time dependence of the relevant mode
of signal after crossing the crystal, but we should take account of the
field dependence on position including a factor $\exp \left( i\mathbf{k}%
_{s}\cdot \mathbf{r}\right) ,$ that is a phase depending on the path length.
Therefore the correct form of eq.$\left( \ref{42}\right) $ would be, modulo
a global phase, 
\begin{equation}
a_{s}(\mathbf{r,}t)=[\left( 1+\frac{1}{2}\left| C\right| ^{2}\right)
a_{s}(0)-iCa_{i}^{*}\left( 0\right) ]\exp \left( i\mathbf{k}_{s}\mathbf{%
\cdot r}-i\omega _{s}t\right) .  \label{44}
\end{equation}
A similar result is obtained for $a_{i}\left( t\right) ,$ that is 
\begin{equation}
a_{i}(\mathbf{r,}t)=[\left( 1+\frac{1}{2}\left| C\right| ^{2}\right)
a_{i}(0)-iCa_{s}^{*}\left( 0\right) ]\exp \left( i\mathbf{k}_{i}\mathbf{%
\cdot r}-i\omega _{i}t\right) .  \label{43}
\end{equation}
Eq.$\left( \ref{44}\right) $ may be interpreted saying that the vacuum
signal is modified by the addition of an amplification of the vacuum idler,
but it travels in the same direction of the incoming vacuum signal, and
therefore it has sense adding the initial vacuum signal plus the
amplification of the idler. And similarly for $a_{i}$ eq.$\left( \ref{43}%
\right) .$ We may perform a change from $C$ to the new parameter $D=\left( 1+%
\frac{1}{2}\left| C\right| ^{2}\right) ^{-1}C$, whence eqs.$\left( \ref{b2}%
\right) $ become, to order $O\left( \left| D\right| ^{2}\right) $, 
\begin{eqnarray}
E_{s}^{+} &=&\left( 1+\frac{1}{2}\left| C\right| ^{2}\right) \left[
a_{s}+Da_{i}^{*}\right] \exp \left( i\mathbf{k}_{s}\mathbf{\cdot r}-i\omega
_{s}t\right) ,  \nonumber \\
E_{i}^{+} &=&\left( 1+\frac{1}{2}\left| C\right| ^{2}\right) \left[
a_{i}+Da_{s}^{*}\right] \exp \left( i\mathbf{k}_{i}\mathbf{\cdot r}-i\omega
_{i}t\right) ,\left| D\right| <<1,  \label{b0}
\end{eqnarray}
and I will ignore the constant global factor $\left( 1+\frac{1}{2}\left|
C\right| ^{2}\right) \sim 1$ because we will be interested in calculating 
\textit{relative} detection rates.

Still eqs.$\left( \ref{44}\right) $ and $\left( \ref{43}\right) ,$ although
good enough for calculations are bad representations of the physics. In fact
a physical beam corresponds to a superposition of the amplitudes, $a_{%
\mathbf{k}}^{*},$ of many modes with frequencies and wavevectors close to $%
\omega _{s}$ and $\mathbf{k}_{s},$ respectively. For instance we may
represent the positive frequency part of the idler beam created in the
crystal, at first order in $D $, as follows 
\begin{equation}
E_{i}^{\left( +\right) }\left( \mathbf{r},t\right) =-iD\int f_{i}\left( 
\mathbf{k}\right) d^{3}\mathbf{k}a_{\mathbf{k}}^{*}\exp \left[ i\left( 
\mathbf{k}-\mathbf{k}_{s}\right) \mathbf{\cdot r}-i\left( \omega -\omega
_{s}\right) t\right] +E_{ZPF}^{\left( +\right) },  \label{45}
\end{equation}
where $\omega =\omega \left( \mathbf{k}\right) $ and $f_{i}\left( \mathbf{k}%
\right) $ is an appropriate function, with domain in a region of $\mathbf{k}$
around $\mathbf{k}_{s}.$ The field $E_{ZPF}^{\left( +\right) }$ is the sum
of amplitudes of all vacuum modes, including the one represented by $a_{s}$
in eq.$\left( \ref{44}\right) .$ (We have neglected a term of order $\left|
C\right| ^{2}$ so that $E_{i}^{\left( +\right) }$ is correct to order $%
\left| C\right| )$. These vacuum modes have fluctuating amplitudes with a
probability distribution given by the vacuum Wigner function eq.$\left( \ref
{1}\right) .$ It may appear that the amplitude $a_{s}$ is lost `as a needle
in the haystack' within the background of many radiation modes, but it is
relevant in photon correlation experiments. In fact the vacuum amplitude $%
a_{s}$ in eqs.$\left( \ref{42}\right) $ or $\left( \ref{44}\right) $ is
fluctuating and the same fluctuations appear also in the signal amplitude $%
a_{s}^{*}$ of eq.$\left( \ref{43}\right) $. Therefore coincidence counts
will be favoured when large positive fluctuations of the fields eqs.$\left( 
\ref{42}\right) $ and $\left( \ref{43}\right) $ arrive simultaneously to
Alice and Bob detectors. In the Hilbert-space formamism this fact is named
`entanglement between a signal and the vacuum'. In the WW formalism of
quantum optics \textit{entanglement appears as a correlation between fields
in distant places}, \textit{including the vacuum fields.}

The mentioned properties of the WW formalism are sufficient for the
interpretation of experiments involving pure radiation field interacting
with macroscopic bodies, these defined by their bulk electric properties
like the refraction index or the nonlinear electrical susceptibility. Within
the WW formalism the interaction between the fields (either signals or
vacuum fields) and macroscopic bodies may be treated as in classical
electrodynamics. This is for instance the case for the action of a laser on
a crystal with nonlinear susceptibility, studied elsewhere\cite{Dechoum}, 
\cite{MS}.

\section{ Photon pairs entangled in polarization}

In this section I will apply the WW formalism to the description of the
polarization correlation of entangled photon pairs produced via spontaneous
parametric down-conversion (SPDC). I will assume that the experimental
set-up is made so that the fields arriving at the detectors correspond to
photon pairs maximally entangled in polarization. These fields are obtained
in the outgoing channels of a beam splitter after sending the signal and
idler beams produced by SPDC to the incoming channels. The electromagnetic
radiation is a vector field with two possible polarizations and I should
take into account this fact including vectors in the description. Thus I
will write the beams produced as follows 
\begin{eqnarray}
\mathbf{E}_{A}^{+} &=&\left( a_{s}+Da_{i}^{*}\right) \exp \left( -i\omega
_{s}t\right) \mathbf{v}+i\left( a_{i}+Da_{s}^{*}\right) \exp \left( -i\omega
_{i}t\right) \mathbf{h,}  \nonumber \\
\mathbf{E}_{B}^{+} &=&-i\left( a_{s}+Da_{i}^{*}\right) \exp \left( -i\omega
_{s}t\right) \mathbf{h}+\left( a_{i}+Da_{s}^{*}\right) \exp \left( -i\omega
_{i}t\right) \mathbf{v,}  \label{0b}
\end{eqnarray}
where $\mathbf{h}$ is a unit vector horizontal and $\mathbf{v}$ vertical. We
have not written explicitly the dependence on position, that could be
restored without difficulty, see eq.$\left( \ref{45}\right) $. Furthermore
from now on I will ignore all spacetime dependence that usually contributes
phase factors irrelevant for our argument. The complex conjugate of the
above fields will be labelled as follows 
\[
(\mathbf{E}_{A}^{+})^{*}\equiv \mathbf{E}_{A}^{-},(\mathbf{E}%
_{B}^{+})^{*}\equiv \mathbf{E}_{B}^{-} 
\]

Eqs.$\left( \ref{0b}\right) $ represent ``two photons entangled in
polarization'' as seen in the Weyl-Wigner formalism. The beams will arrive
at the Alice and Bob polarization analyzers put at angles $\theta $ and $%
\phi $ with the vertical respectively. Hence the beams emerging from them
will have field amplitudes

\begin{eqnarray}
E_{A}^{+} &=&\left( a_{s}+Da_{i}^{*}\right) \cos \theta +i\left(
a_{i}+Da_{s}^{*}\right) \sin \theta ,  \nonumber \\
E_{B}^{+} &=&-i\left( a_{s}+Da_{i}^{*}\right) \sin \phi +\left(
a_{i}+Da_{s}^{*}\right) \cos \phi ,  \label{b3}
\end{eqnarray}
and polarizations at angles $\theta $ and $\phi $ with the vertical,
respectively. For later convenience I define the partial fields

\begin{eqnarray}
E_{A0}^{+} &=&a_{s}\cos \theta +ia_{i}\sin \theta ,E_{B0}^{+}=-ia_{s}\sin
\phi +a_{i}\cos \phi ,  \label{b4} \\
E_{A1}^{+} &=&D\left[ a_{i}^{*}\cos \theta +ia_{s}^{*}\sin \theta \right]
,E_{B1}^{+}=D\left[ -ia_{i}^{*}\sin \phi +a_{s}^{*}\cos \phi \right] . 
\nonumber
\end{eqnarray}
Hence we may define intensities as follows

\begin{eqnarray}
I_{A0}
&=&E_{A0}^{+}E_{A0}^{-},I_{A1}=E_{A0}^{+}E_{A1}^{-}+E_{A1}^{+}E_{A0}^{-}+E_{A1}^{+}E_{A1}^{-},
\nonumber \\
I_{B0}
&=&E_{B0}^{+}E_{B0}^{-},I_{B1}=E_{B0}^{+}E_{B1}^{-}+E_{B1}^{+}E_{B0}^{-}+E_{B1}^{+}E_{B1}^{-},
\nonumber \\
I_{A}
&=&E_{A}^{+}E_{A}^{-}=I_{A0}+I_{A1},I_{B}=E_{B}^{+}E_{B}^{-}=I_{B0}+I_{B1}.
\label{b5}
\end{eqnarray}

The single, $P_{A}$, $P_{B}$, and coincidence, $P_{AB}$, detection rates in
the WW formalism may be obtained by comparison with the rates calculated in
the Hilbert-space formalism. Thus in the following we revisit briefly the
Hilbert-space treatment of the entangled photon pairs. I will start with the
quantum fields arriving at Alice and Bob respectively, that are the
counterparts of the WW eqs.$\left( \ref{b3}\right) .$ It is trivial to get
them either from eq.$\left( \ref{40}\right) $ or, taking eqs.$\left( \ref{2}%
\right) $ into account, that is ``putting hats'' in the WW eqs.$\left( \ref
{b4}\right) .$ We get the field operators 
\[
\hat{E}_{A}^{+}=\hat{E}_{A0}^{+}+\hat{E}_{A1}^{+},\hat{E}_{B}^{+}=\hat{E}%
_{B0}^{+}+\hat{E}_{B1}^{+}, 
\]
\begin{eqnarray}
\hat{E}_{A0}^{+} &=&\hat{a}_{s}\cos \theta +i\hat{a}_{i}\sin \theta ,\hat{E}%
_{B0}^{+}=-i\hat{a}_{s}\sin \phi +\hat{a}_{i}\cos \phi ,  \nonumber \\
\hat{E}_{A1}^{+} &=&D[\hat{a}_{i}^{\dagger }\cos \theta +i\hat{a}%
_{s}^{\dagger }\sin \theta ],\hat{E}_{B1}^{+}=D[-i\hat{a}_{i}^{\dagger }\sin
\phi +\hat{a}_{s}^{\dagger }\cos \phi ],  \label{3b}
\end{eqnarray}
and similar for the Hermitean conjugates. Alice single detection rate is
proportional to the following vacuum expectation (with $\hat{E}%
_{A}^{-}=\left( \hat{E}_{A}^{+}\right) ^{\dagger })$%
\begin{eqnarray}
P_{A} &=&\langle 0\mid \hat{E}_{A}^{-}\hat{E}_{A}^{+}\mid 0\rangle =\left|
D\right| ^{2}\langle 0\mid \hat{E}_{A1}^{-}\hat{E}_{A1}^{+}\mid 0\rangle 
\nonumber \\
&=&\left| D\right| ^{2}\langle 0\mid (\hat{a}_{i}\cos \theta -i\hat{a}%
_{s}\sin \theta )(\hat{a}_{i}^{\dagger }\cos \theta +i\hat{a}_{s}^{\dagger
}\sin \theta )\mid 0\rangle  \nonumber \\
&=&\left| D\right| ^{2}\langle 0\mid \hat{a}_{i}\hat{a}_{i}^{\dagger }\cos
^{2}\theta +\hat{a}_{s}\hat{a}_{s}^{\dagger }\sin ^{2}\theta \mid 0\rangle
=\left| D\right| ^{2},  \label{4b}
\end{eqnarray}
where in the former equality I have neglected creation $\hat{E}_{A0}^{-}$
(annihilation $\hat{E}_{A0}^{+}$) operators appearing on the left (right). A
similar result may be obtained for the single detection rate of Bob, that is 
\begin{equation}
P_{B}=\langle 0\mid \hat{E}_{B}^{-}\hat{E}_{B}^{+}\mid 0\rangle =\left|
D\right| ^{2},\hat{E}_{B}^{-}=\left( \hat{E}_{B}^{+}\right) ^{\dagger }.
\label{4e}
\end{equation}
We are assuming ideal detectors, but for real detectors $P_{A}$ and $P_{B}$
should be multiplied times the detection efficiencies $\eta _{A}$ and $\eta
_{B},$ and the coincidence rate $P_{AB}$ times $\eta _{A}\eta _{B}.$

In order to get the detection rule for single rates in the WW formalism%
\textit{\ }we should translate eq.$\left( \ref{4b}\right) $ taking eqs.$%
\left( \ref{3a}\right) $ into account. We get 
\begin{eqnarray}
P_{A} &=&\left[ (\left\langle \left| a_{i}\right| ^{2}\right\rangle -\frac{1%
}{2})\cos ^{2}\theta +(\left\langle \left| a_{s}\right| ^{2}\right\rangle -%
\frac{1}{2})\sin ^{2}\theta \right]  \nonumber \\
&&+\left| D\right| ^{2}\left[ (\left\langle \left| a_{s}\right|
^{2}\right\rangle +\frac{1}{2})\cos ^{2}\theta +(\left\langle \left|
a_{i}\right| ^{2}\right\rangle +\frac{1}{2})\sin ^{2}\theta \right] 
\nonumber \\
&=&\left| D\right| ^{2}\left[ \cos ^{2}\theta +\sin ^{2}\theta \right]
=\left| D\right| ^{2},\left\langle a_{i}a_{s}^{*}\right\rangle =\left\langle
a_{s}a_{i}^{*}\right\rangle =0,  \label{PW}
\end{eqnarray}
that agrees with the result calculated in the Hilbert-space formalism, eq.$%
\left( \ref{4d}\right) ,$ as it should. Now we compare eq.$\left( \ref{PW}%
\right) $ with the average of the field intensity arriving at Alice (see eq.$%
\left( \ref{b3}\right) )$, that is 
\begin{eqnarray}
I_{A} &=&\left| E_{A}^{+}\right| ^{2},\left\langle I_{A}\right\rangle
=\left\langle \left| a_{i}\right| ^{2}\right\rangle \cos ^{2}\theta
+\left\langle \left| a_{s}\right| ^{2}\right\rangle \sin ^{2}\theta 
\nonumber \\
&&+\left| D\right| ^{2}\left[ \left\langle \left| a_{s}\right|
^{2}\right\rangle \cos ^{2}\theta +\left\langle \left| a_{i}\right|
^{2}\right\rangle \sin ^{2}\theta \right] =\frac{1}{2}\left( 1+\left|
D\right| ^{2}\right) .  \label{IA}
\end{eqnarray}
We see that going from eq.$\left( \ref{IA}\right) $ to eq.$\left( \ref{PW}%
\right) $ the signal terms (those of order $\left| D\right| ^{2})$ are
multiplied times 2, whilst those coming from the vacuum (of order unity) are
eliminated. This may be seen as a subtraction of the vacuum (ZPF) and
multiplication of the signal times 2, which leads to the following rule for
the single detection rate in the WW formalism: 
\begin{equation}
P_{A}=2\left\langle I_{A}\right\rangle -2\left\langle I_{A0}\right\rangle
,P_{B}=2\left\langle I_{B}\right\rangle -2\left\langle I_{B0}\right\rangle ,
\label{b6}
\end{equation}
the latter for Bob detection rate.

The Hilbert-space rule for the coincidence rate is the vacuum expectation
value of the product of four field operators in normal order. Here we have
two terms 
\begin{equation}
P_{AB}=\frac{1}{2}\langle 0\mid \hat{E}_{A}^{-}\hat{E}_{B}^{-}\hat{E}_{B}^{+}%
\hat{E}_{A}^{+}\mid 0\rangle +\frac{1}{2}\langle 0\mid \hat{E}_{B}^{-}\hat{E}%
_{A}^{-}\hat{E}_{A}^{+}\hat{E}_{B}^{+}\mid 0\rangle ,  \label{expect}
\end{equation}
that would be equal if $\hat{E}_{A}^{+}$ and $\hat{E}_{B}^{+}$ commute. The
former expectation may be evaluated to order $\left| D\right| ^{2}$ as
follows 
\begin{eqnarray*}
\langle 0 &\mid &\hat{E}_{A}^{-}\hat{E}_{B}^{-}\hat{E}_{B}^{+}\hat{E}%
_{A}^{+}\mid 0\rangle =\langle 0\mid \hat{E}_{A1}^{-}\hat{E}_{B}^{-}\hat{E}%
_{B}^{+}\hat{E}_{A1}^{+}\mid 0\rangle =\langle 0\mid \hat{E}_{A1}^{-}\hat{E}%
_{B0}^{-}\hat{E}_{B0}^{+}\hat{E}_{A1}^{+}\mid 0\rangle \\
&=&\langle 0\mid \hat{E}_{A1}^{-}\hat{E}_{B0}^{-}\mid 0\rangle \langle 0\mid 
\hat{E}_{B0}^{+}\hat{E}_{A1}^{+}\mid 0\rangle =\left| \langle 0\mid \hat{E}%
_{B0}^{+}\hat{E}_{A1}^{+}\mid 0\rangle \right| ^{2} \\
&=&\left| D\right| ^{2}\left| \langle 0\mid \left( -i\hat{a}_{s}\sin \phi +%
\hat{a}_{i}\cos \phi \right) \left( \hat{a}_{i}^{\dagger }\cos \theta +i\hat{%
a}_{s}^{\dagger }\sin \theta \right) \mid 0\rangle \right| ^{2},
\end{eqnarray*}
where the former equality, similar to eq.$\left( \ref{4b}\right) ,$ removes
creation operators on the left and annihilation operators on the right, the
second one removes terms of order $\left| D\right| ^{4}$ and the rest is
trivial. The latter term of eq.$\left( \ref{expect}\right) $ gives a similar
contribution so that we get 
\begin{equation}
P_{AB}=\frac{1}{2}\left| \langle 0\mid \hat{E}_{B0}^{+}\hat{E}_{A1}^{+}\mid
0\rangle \right| ^{2}+\frac{1}{2}\left| \langle 0\mid \hat{E}_{A0}^{+}\hat{E}%
_{B1}^{+}\mid 0\rangle \right| ^{2}=\left| D\right| ^{2}\cos ^{2}(\theta
-\phi ).  \label{4g}
\end{equation}
Here the creation operators are placed to the right and those of
annihilation to the left, so that no subtraction is required in passing to
the WW formalism. It is enough to substitute c-number amplitudes mutiplied
times 2 for the field operators, in order to get the rule for the
coincidence rate in the WW formalism. That is

\begin{eqnarray}
\left\langle E_{A0}^{+}E_{B1}^{+}\right\rangle +\left\langle
E_{A1}^{+}E_{B0}^{+}\right\rangle &=&\left\langle
E_{A}^{+}E_{B}^{+}\right\rangle =D\cos (\theta -\phi )  \nonumber \\
\left| \left\langle E_{A}^{+}E_{B}^{+}\right\rangle \right| ^{2} &=&\left|
D\right| ^{2}\cos ^{2}(\theta -\phi ),  \label{9b}
\end{eqnarray}
where we have taken eqs.$\left( \ref{b4}\right) $ and $\left( \ref{3a}%
\right) $ into account. Here the vacuum subtraction is not explicit because
the vacuum term would be zero, that is $\left\langle
E_{A0}^{+}E_{B0}^{+}\right\rangle =0.$

It is interesting to get the coincidence detection rate in terms of field
intensities, rather than amplitudes. To do that we start calculating

\begin{equation}
\left\langle I_{A}I_{B}\right\rangle =\left\langle
E_{A}^{+}E_{A}^{-}E_{B}^{+}E_{B}^{-}\right\rangle .  \label{b7}
\end{equation}
In the WW formalism the field amplitudes are c-numbers, therefore they
commute, and the averages should be performed as in eq.$\left( \ref{3a}%
\right) .$ The expectation eq.$\left( \ref{b8}\right) $ may be obtained
taking into account that the fields have the mathematical properties of
Gaussian random variables, see eq.$\left( \ref{1}\right) $ (although this
section is devoted to calculations and for the moment I am not commited to
any physical interpretation). Thus I apply a well known property of the
average of the product of four Gaussian random variables, that is

\begin{eqnarray}
\left\langle I_{A}I_{B}\right\rangle &=&\left\langle
E_{A}^{+}E_{A}^{-}\right\rangle \left\langle E_{B}^{+}E_{B}^{-}\right\rangle
+\left\langle E_{A}^{+}E_{B}^{-}\right\rangle \left\langle
E_{A}^{-}E_{B}^{+}\right\rangle +\left\langle
E_{A}^{+}E_{B}^{+}\right\rangle \left\langle E_{A}^{-}E_{B}^{-}\right\rangle
\nonumber \\
&=&\left\langle I_{A}\right\rangle \left\langle I_{B}\right\rangle +\left| \
\langle E_{A}^{+}E_{B}^{-}\rangle \right| ^{2}+\left| \left\langle
E_{A}^{+}E_{B}^{+}\right\rangle \right| ^{2}.  \label{b9}
\end{eqnarray}
A similar procedure but involving the vacuum intensities, gives 
\begin{equation}
\left\langle I_{A0}I_{B0}\right\rangle =\left\langle I_{A0}\right\rangle
\left\langle I_{B0}\right\rangle +\left| \left\langle
E_{A0}^{+}E_{B0}^{-}\right\rangle \right| ^{2}+\left| \left\langle
E_{A0}^{+}E_{B0}^{+}\right\rangle \right| ^{2}.  \label{10b}
\end{equation}
Here the third term does not contribute and the second one equals the second
term of eq.$\left( \ref{b9}\right) $ to order $\left| D\right| ^{2}.$ Hence
we get the rule for the coincidence rate in the WW formalism that I write
both in terms of fields and in terms of intensities as follows 
\begin{equation}
P_{AB}=\left| \left\langle E_{A}^{+}E_{B}^{+}\right\rangle \right|
^{2}=\left\langle I_{A}I_{B}\right\rangle -\left\langle I_{A}\right\rangle
\left\langle I_{B}\right\rangle -\left\langle I_{A0}I_{B0}\right\rangle
+\left\langle I_{A0}\right\rangle \left\langle I_{B0}\right\rangle .
\label{b11}
\end{equation}

\section{Locality in the experiments with entangled photon pairs}

\subsection{A realistic interpretation of the experiments}

I emphasize again that the WW formalism provides an alternative formulation
of quantum optics, physically equivalent to the more common the
Hilbert-space formalism. But it suggests a picture of the optical phenomena
quite different from the latter. Indeed the picture may provide a local
realistic interpretation in terms of random variables and stochastic
processes. In the following I present the main ideas of this stochastic
interpretation. It rests upon several assumptions as follows.

The fundamental hypothesis is that \textit{the} \textit{electromagnetic
vacuum field is a real stochastic field }(\textit{the zeropoint field, ZPF)}%
. If expanded in normal modes the ZPF has a (positive) probability
distribution of the amplitudes given by eq.$\left( \ref{1}\right) .$
Therefore we assume that all bodies are immersed in ZPF absorbing and
emitting radiation continuously. In particular the ground state of an atom
corresponds to a dynamical equilibrium with the ZPF. The hypothesis that
vacuum fields are real stochastic fields leads to a general interpretation
of quantum theory that has been sketched elsewhere\cite{FOS}.

According to that assumption any photodetector would be immersed in an
extremely strong radiation, infinite if no cut-off existed. Thus how might
we explain that detectors are not activated by the vacuum radiation? Firstly
the strong vacuum field is effectively reduced to a weaker level if \textit{%
we assume that only radiation within some (small) frequency interval is able
to activate a photodetector}, that is the interval of sensibility $\left(
\omega _{1},\omega _{2}\right) $. However the problem is not yet solved
because signals involved in experiments have typical intensities of order
the vacuum radiation in the said frequency interval so that the detector
would be unable to distinguish a signal from the ZPF noise. Our proposal is
to assume that \textit{a detector may be activated only when the net
Poynting vector (i. e. the directional energy flux) of the incoming
radiation is different from zero, including both signal and vacuum fields}.
More specifically I will assume that \textit{the detector possesses an
active area and the probability of a photocount is proportional to the net
radiant energy flux crossing that area from the front side during the
activation time, T},\textit{\ the probability being zero if the net flux
crosing the area is in the reverse direction.}

These assumptions allow to understand qualitatively why the signals, but not
the vacuum fields, activate detectors. Indeed the ZPF arriving at any point
(in particular the detector) would be isotropic on the average, whence its
associated mean Poynting vector would be nil, therefore only the signal
radiation should produce photocounts. A problem remains because the vacuum
fields are fluctuating so that the net Poynting vector also fluctuates.
These problems diminish due to the fact that \textit{photocounts are not
produced by an instantaneous interaction of the fields with the detectors
but the activation requires some time interval}, a known fact in
experiments. This would reduce the effect of the vacuum fluctuations.

Our aim is to achieve a realistic local interpretation of the experiments
measuring polarization correlation of entangled photon pairs, that we
studied with the WW formalism in the previous section. Thus I will consider
two vacuum beams entering the nonlinear crystal, where they give rise to a
``signal'' and an ``idler'' beams. After crossing several appropriate
devices they produce fields that will arrive at the Alice and Bob detectors.
I do not attempt to present a detailed model that should involve many modes
in order to represent the signals as (narrow) beams, see eq.$\left( \ref{45}%
\right) .$

In agreement with our previous hypotheses a photodetection should derive
from the net energy flux crossing the active photocounter surface. Thus I
will assume that the detection probabilities per time window, that is the
single $P_{A},P_{B}$ and coincidence $P_{AB}$ detection rates, are 
\begin{eqnarray}
P_{A} &=&\left\langle \left[ M_{A}\right] _{+}\right\rangle
,P_{B}=\left\langle \left[ M_{A}\right] _{+}\right\rangle
,P_{AB}=\left\langle \left[ M_{A}\right] _{+}\left[ M_{B}\right]
_{+}\right\rangle ,  \nonumber \\
M_{A} &\equiv &T^{-1}\int_{0}^{T}\vec{n}_{A}\cdot \vec{I}_{total}^{A}\left( 
\mathbf{r}_{A}\mathbf{,}t\right) dt,  \nonumber \\
M_{B} &\equiv &T^{-1}\int_{0}^{T}\vec{n}_{B}\cdot \vec{I}_{total}^{B}\left( 
\mathbf{r}_{A},t\right) dt,  \label{s1}
\end{eqnarray}
where $\left[ M\right] _{+}=$ $M$ if $M>0,$ $\left[ M\right] _{+}=0$
otherwise, and $\vec{n}$ is a unit vector in the direction of the incoming
signal beams, that I assume perpendicular to the active area of the
detector. I use units such that both the intensities and the detection rates
are dimensionless (as explained above the rates are defined as probabilities
of detection within a detection time window $T$). In eq.$\left( \ref{s1}%
\right) $ the positivity constraint (i.e. putting $\left[ M_{A}\right] _{+}$
rather than $M_{A})$ is needed because the detection probabilities must be
non-negative for any particular run of an experiment whilst the quantities $%
M_{A}$ and $M_{B}$ are fluctuating and may be negative sometimes.
Nevertheless the ensemble averages involved in eq.$\left( \ref{s1}\right) $
are positive or zero and the fluctuations will not be too relevant due to
the time integration that washes them out. Therefore I will ignore the
positivity constraint substituting $M_{A}$ for $\left[ M_{A}\right] _{+}$
and $M_{B}$ for $\left[ M_{B}\right] _{+}$ in the following.

In order to have a realistic model of the experiments I will consider a
simplified treatement involving just two vacuum radiation modes, with
amplitudes $a_{s}$ and $a_{i},$ as in the WW calculation of the previous
section. After crossing several appropriate devices the fields will arrive
at the Alice and Bob detectors. Each one of these two fields consists of two
parts, one of order $0$ and another of order $\left| D\right| <<1.$ The
former, given in eqs.$\left( \ref{b4}\right) ,$ is what would arrive at the
detectors if there was no pumping laser and therefore no signal. It is just
a part of the ZPF, whilst the rest of the ZPF consists of radiation not
appearing in the equations of the previous section because they were not
needed in the calculations. The total ZPF should have the property of
isotropy, therefore giving nil net flux in the detector (modulo fluctuations
that we shall ignore at that moment). The terms of order $\left| D\right| ,$
given by eqs.$\left( \ref{b4}\right) $ correspond to the signal produced in
the nonlinear crystal after the modifications introduced by beam splitters
and polarizers (and other devices like apertures, filters, lens systems,
etc. whose action is not detailed in our simplified model). In summary the
Poynting vectors of the radiation at the (center of the) active area of the
detectors may be written 
\begin{eqnarray}
Alice &:&\vec{I}_{total}^{A}\left( t\right) =\vec{I}_{ZPF}^{A}\left(
t\right) +\vec{I}_{A}\left( t\right) ,  \nonumber \\
Bob &:&\vec{I}_{total}^{B}\left( t\right) =\vec{I}_{ZPF}^{B}\left( t\right) +%
\vec{I}_{B}\left( t\right) .\smallskip  \label{s2}
\end{eqnarray}
$\vec{I}_{A}$, $\vec{I}_{B},$ are due to the fields $E_{A}$, $E_{B}$, eqs.$%
\left( \ref{b3}\right) $, coming from the fields emerging from the
non-linear crystal. The Poynting vectors $\vec{I}_{A}\left( t\right) $ and $%
\vec{I}_{B}\left( t\right) $ have the direction of $\vec{n}_{A}$ and $\vec{n}%
_{B\text{ }}$respectively, see eqs.$\left( \ref{s1}\right) ,$ and their
moduli would be the field intensities. Furthermore these intensities are
time independent (in our representation of the fields), so that we may write 
\begin{equation}
M_{A}=I_{ZPF}^{A}+I_{A},I_{ZPF}^{A}\equiv T^{-1}\int_{0}^{T}\vec{n}_{A}\cdot 
\vec{I}_{ZPF}^{A}\left( \mathbf{r}_{A}\mathbf{,}t\right) dt  \label{s3}
\end{equation}
and similar for $M_{B}.$ In order to get the Alice single detection rate we
need the average of $M_{A},$ that we will evaluate by comparison with the
case when there is no pumping on the nonlinear crystal. In this case $I_{A}$
becomes the intensity $I_{A0}$ and the Poynting vector of all vacuum fields
arriving at the detector of Alice, i. e. $\vec{I}_{ZPF}^{A}\left( t\right) +%
\vec{I}_{A0}\left( t\right) ,$ should have nil average due to the isotropy
of the $ZPF$. And similar for Bob. As a consequence the intensities $I_{A0}$
and $I_{B0},$ eqs.$\left( \ref{b5}\right) ,$ should fulfil the following
equalities 
\begin{equation}
\left\langle I_{ZPF}^{A}+I_{A0}\right\rangle =\left\langle
I_{ZPF}^{B}+I_{B0}\right\rangle =0.  \label{s4}
\end{equation}
It would appear that this relation could not be true for all values of the
angles $\theta ,\phi $ eqs.$\left( \ref{b4}\right) $ because the ZPF
Poynting vector $\hat{I}_{ZPF}^{A}$ and $\hat{I}_{ZPF}^{B}$ should not
depend on our choice of angles whilst $I_{A0}$ and $I_{B0}$ do depend.
However the positions of the polarizers may influence also the ZPZ arriving
at the detectors and it is plausible that the total Poynting vector has
always zero mean. From eqs.$\left( \ref{s3}\right) $ and $\left( \ref{s4}%
\right) $ we may derive the single rates of Alice and Bob, that is 
\begin{equation}
P_{A}=\left\langle M_{A}\right\rangle =\left\langle I_{A}\right\rangle
-\left\langle I_{A0}\right\rangle =\frac{1}{2}\left| D\right|
^{2},P_{B}=\left\langle M_{B}\right\rangle =\frac{1}{2}\left| D\right| ^{2}.
\label{s5}
\end{equation}

The coincidence detection rate may be got taking eq.$\left( \ref{s1}\right) $
into account, that is 
\begin{equation}
P_{AB}=\left\langle M_{A}M_{B}\right\rangle =\langle \left[
I_{ZPF}^{A}+I_{A}\right] \left[ I_{ZPF}^{B}+I_{B}\right] \rangle .
\label{s6}
\end{equation}
For the sake of clarity the calculation of this average is made in some
detail as follows. I will start deriving the joint probability distribution
of $M_{A}$ and $M_{B}$ from the distributions of the amplitudes of normal
modes, eq.$\left( \ref{1}\right) $, that is 
\begin{equation}
\rho \left( M_{A},M_{B}\right) =\int W_{0}\prod_{j}da_{j}da_{j}^{*}\delta
\left( M_{A}-I_{ZPF}^{A}-I_{A}\right) \delta \left(
M_{B}-I_{ZPF}^{B}-I_{B}\right)  \label{s9}
\end{equation}
where $\delta \left( {}\right) $ is Dirac\'{}s delta and $I_{A},I_{B}$ are
functions of the two modes involved, i. e. those with amplitudes $a_{s}$, $%
a_{i},$ see eqs.$\left( \ref{b3}\right) $. Hence the expectation eq.$\left( 
\ref{s6}\right) $ becomes

\begin{eqnarray}
P_{AB} &=&\left\langle M_{A}M_{B}\right\rangle =\int M_{A}M_{B}\rho \left(
M_{A},M_{B}\right) dM_{A}dM_{B}  \nonumber \\
&=&\int W_{0}\prod_{j}da_{j}da_{j}^{*}\left[ I_{ZPF}^{A}+I_{A}\right] \left[
I_{ZPF}^{B}+I_{B}\right]  \nonumber \\
&\equiv &\left\langle \left[ I_{ZPF}^{A}+I_{A}\right] \left[
I_{ZPF}^{B}+I_{B}\right] \right\rangle ,  \label{s8}
\end{eqnarray}
where we have performed the $M_{A}$ and $M_{B}$ integrals in the former
equality.

If there was no pumping laser on the nonlinear crystal the joint detection
rate should be zero whence eq.$\left( \ref{s8}\right) $ leads to 
\begin{equation}
\left\langle \left[ I_{ZPF}^{A}+I_{A0}\right] \left[
I_{ZPF}^{B}+I_{B0}\right] \right\rangle =0.  \label{s10}
\end{equation}
From eqs.$\left( \ref{s4}\right) ,\left( \ref{s8}\right) $ and $\left( \ref
{s10}\right) $ it is possible to get $P_{AB}.$ Firstly I point out that $%
I_{ZPF}^{A}$ and $I_{ZPF}^{B}$ could not depend on whether the pumping is
switched on or out. More specifically the probability of the possible values
of $I_{ZPF}$ would be the same whether the pumping is on or out. Then
subtracting eq.$\left( \ref{s10}\right) $ from eq.$\left( \ref{s8}\right) $
we get 
\begin{eqnarray*}
P_{AB} &=&\left\langle \left( I_{B}-I_{B0}\right) I_{ZPF}^{A}\right\rangle
+\left\langle \left( I_{A}-I_{A0}\right) I_{ZPF}^{B}\right\rangle
+\left\langle I_{A}I_{B}\right\rangle -\left\langle I_{A0}I_{B0}\right\rangle
\\
&=&\left\langle I_{B1}I_{ZPF}^{A}\right\rangle +\left\langle
I_{A1}I_{ZPF}^{B}\right\rangle +\left\langle I_{A}I_{B}\right\rangle
-\left\langle I_{A0}I_{B0}\right\rangle ,
\end{eqnarray*}
where we have taken into account the definition of $I_{A1}$ and $I_{B1},$ eq.%
$\left( \ref{b5}\right) $. The ZPF intensity $I_{ZPF}^{A}$ should be
independent of whether $I_{B1}$ is zero (pumping out) or finite (pumping on)
as commented above, therefore $I_{ZPF}^{A}$ and $I_{B1}$ are uncorrelated.
Similarly for $I_{ZPF}^{B}$ and $I_{A1}.$ Hence we may write, taking eq.$%
\left( \ref{s4}\right) $ into account, 
\begin{eqnarray}
P_{AB} &=&\left\langle I_{B1}\right\rangle \left\langle
I_{ZPF}^{A}\right\rangle +\left\langle I_{A1}\right\rangle \left\langle
I_{ZPF}^{B}\right\rangle +\left\langle I_{A}I_{B}\right\rangle -\left\langle
I_{A0}I_{B0}\right\rangle  \nonumber \\
&=&\left\langle I_{A}I_{B}\right\rangle -\left\langle I_{A1}\right\rangle
\left\langle I_{B0}\right\rangle -\left\langle I_{B1}\right\rangle
\left\langle I_{A0}\right\rangle -\left\langle I_{A0}I_{B0}\right\rangle 
\nonumber \\
&=&\left\langle I_{A0}I_{B1}\right\rangle +\left\langle
I_{A1}I_{B0}\right\rangle -\left\langle I_{A1}\right\rangle \left\langle
I_{B0}\right\rangle -\left\langle I_{B1}\right\rangle \left\langle
I_{A0}\right\rangle .  \label{s11}
\end{eqnarray}
Now we may write $\left\langle I_{A0}I_{B1}\right\rangle +\left\langle
I_{A1}I_{B0}\right\rangle $ in terms of the fields taking into account the
property of Gaussian random variables used in the evaluation of eq.$\left( 
\ref{b7}\right) ,$ that is 
\begin{eqnarray*}
\left\langle I_{A0}I_{B1}\right\rangle &=&\left\langle
E_{A0}^{+}E_{A0}^{-}\right\rangle \left\langle
E_{B1}^{+}E_{B1}^{-}\right\rangle +\left\langle
E_{A0}^{+}E_{B1}^{-}\right\rangle \left\langle
E_{A0}^{-}E_{B1}^{+}\right\rangle +\left\langle
E_{A0}^{+}E_{B1}^{+}\right\rangle \left\langle
E_{A0}^{-}E_{B1}^{-}\right\rangle \\
&=&\left\langle I_{A0}\right\rangle \left\langle I_{B1}\right\rangle +\left|
\ \langle E_{A0}^{+}E_{B1}^{-}\rangle \right| ^{2}+\left| \left\langle
E_{A0}^{+}E_{B1}^{+}\right\rangle \right| ^{2},
\end{eqnarray*}
and similar for $\left\langle I_{A1}I_{B0}\right\rangle .$ Then eq.$\left( 
\ref{s11}\right) $ becomes 
\[
P_{AB}=\left| \left\langle E_{A0}^{+}E_{B1}^{+}\right\rangle \right|
^{2}+\left| \left\langle E_{A1}^{+}E_{B0}^{+}\right\rangle \right|
^{2}+\left| \left\langle E_{A0}^{+}E_{B1}^{-}\right\rangle \right|
^{2}+\left| \left\langle E_{A1}^{+}E_{B0}^{-}\right\rangle \right| ^{2}. 
\]
The two latter terms are nil and the former two give, taking eqs.$\left( \ref
{b4}\right) $ into account, 
\begin{equation}
P_{AB}=\frac{1}{2}\left| D\right| ^{2}\cos ^{2}(\theta -\phi ).  \label{s12}
\end{equation}
As a conclusion the results of our realistic model, eqs.$\left( \ref{s5}%
\right) $ and $\left( \ref{s12}\right) ,$ agree with the quantum predictions
eqs.$\left( \ref{4b}\right) ,\left( \ref{4e}\right) $ and $\left( \ref{4g}%
\right) ,$ modulo an scaling parameter 1/2.

Our model, resting upon the WW formalism of quantum optics, provides a
picture quite different from the one suggested by the Hilbert-space in terms
of photons. We do not assume that the photocount probability within a time
window is the product of a \textit{small} probability, of order $\left|
D\right| ^{2},$ that the pumping laser produces in the crystal an
``entangled photon pair'' times a detection probability of order\textit{\
unity} conditional to the photon production. The latter probability is
defined as the detection efficiency (neglecting small losses). In our model
the probabilities of photocounts do not factorize that way. Furthermore the
concept of photon does not appear at all, but there are \textit{continuous
fluctuating fields including a real ZPF} arriving at the detectors that are
activated when the fluctuations are big enough.

It is interesting to study more closely the ``quantum correlation''
qualified as strange from a classical point of view because it is a
consequence of the phenomenon of entanglement. The origin is the correlation
between the signal $I_{B1}$ produced in the crystal and the part $I_{A0}$ of
the ZPF, that is essential for the large value of the coincidence rate eq.$%
\left( \ref{s11}\right) .$ It is remarkable that this correlation derives
from the fact that the same normal mode appears in both radiation fields, $%
E_{A0}^{+}$ and $E_{B1}^{+}$ see eq.$\left( \ref{9b}\right) ,$ received by
Alice and Bob respectively. And similarly for $E_{A1}^{+}$ and $E_{B0}^{+}.$
Also with reference to eqs.$\left( \ref{b4}\right) $ and $\left( \ref{b6}%
\right) $ we see that if $I_{A0}-\left\langle I_{A0}\right\rangle $ is
positive for some value of the random variable $I_{A0}$ then $I_{B1}$ is
large, but if $I_{A0}-\left\langle I_{A0}\right\rangle $ is negative then $%
I_{B1}$ is small. This argument explains intuitively why the product is
positive on the average, that is proved quantitatively by eq.$\left( \ref
{s11}\right) $. Furthermore I stress that the ensemble average of $%
I_{A0}-\left\langle I_{A0}\right\rangle $ is zero, meaning that \textit{only
the fluctuations} are involved in the enhancement of detection probability
by Bob due to the intensity $I_{B1}$. And similarly for the fluctuations of
(Bob) term $I_{B0}-\left\langle I_{B0}\right\rangle $ that enhance the
detection probability of Alice due to $I_{A1}$. This leads to an interesting
interpretation of entanglement: \textit{it is a correlation between }\textbf{%
fluctuations}\textit{\ involving the vacuum fields}.

Finally I stress that the hypothesis that the quantum vacuum fields are real
allows a concept of locality weaker than Bell\'{}s. Indeed the signal fields
(accompanied by vacuum fields) travel causally from the source (the laser
pumping beam and the nonlinear crystal folowed by a beam-splitter and other
devices) to the detectors. Thus I claim that the model of this paper is
local. On the other hand the results for the single and coincidence
detection probabilities within a time window, eqs.$\left( \ref{s5}\right) $
and $\left( \ref{s12}\right) ,$ violate a Clauser-Horne inequality $\left( 
\cite{CH}\right) $ as may be easily checked. Therefore it is possible a
definition of locality, weaker than Bell\'{}s, able to explain the entangled
photon experiments without the need of assuming the violation of
relativistic causality.

\end{document}